\documentclass[12pt]{iopart}
\usepackage{graphicx}
\usepackage{epstopdf}
\usepackage{algorithmic}
\usepackage{algorithm}
\usepackage{booktabs} 
\usepackage{xcolor}

\makeatletter
\newenvironment{breakablealgorithm}
{% \begin{breakablealgorithm}
		\begin{center}
			\refstepcounter{algorithm}% New algorithm
			\hrule height.8pt depth0pt \kern2pt% \@fs@pre for \@fs@ruled
			\renewcommand{\caption}[2][\relax]{% Make a new \caption
				{\raggedright\textbf{\ALG@name~\thealgorithm} ##2\par}%
				\ifx\relax##1\relax % #1 is \relax
				\addcontentsline{loa}{algorithm}{\protect\numberline{\thealgorithm}##2}%
				\else % #1 is not \relax
				\addcontentsline{loa}{algorithm}{\protect\numberline{\thealgorithm}##1}%
				\fi
				\kern2pt\hrule\kern2pt
			}
		}{% \end{breakablealgorithm}
		\kern2pt\hrule\relax% \@fs@post for \@fs@ruled
	\end{center}
}
\makeatother

\begin{document}

\title{A Graph-Neural-Network-Entropy model of vital node identification on network attack and propagation}

\author{Huaizhi Liao, Tian Qiu\footnote[7]{Corresponding author. E-mail: tianqiu.edu@gmail.com} and Guang Chen}
\address{School of Information Engineering, Nanchang Hangkong University, Nanchang, 330063, China}
\vspace{10pt}

\begin{abstract}
Vital nodes usually play a key role in complex networks. Uncovering these nodes is an important task in protecting the network, especially when the network suffers intentional attack. Many existing methods have not fully integrated the node feature, interaction and state. In this article, we propose a novel method (GNNE) based on graph neural networks and information entropy. The method employs a Graph Convolutional Network (GCN) to learn the nodes' features, which are input into a Graph Attention Network (GAT) to obtain the influence factor of nodes, and the node influence factors are used to calculate the nodes' entropy to evaluate the node importance. The GNNE takes advantage of the GCN and GAT, with the GCN well extracting the nodes' features and the GAT aggregating the features of the nodes' neighbors by using the attention mechanism to assign different weights to the neighbors with different importance, and the nodes' entropy quantifies the nodes' state in the network. The proposed method is trained on a synthetic Barabási–Albert network, and tested on six real datasets. Compared with eight traditional topology-based methods and four graph-machine-learning-based methods, the GNNE shows an advantage for the vital node identification in the perspectives of network attack and propagation. 
\end{abstract}

\vspace{2pc}
\noindent{\it Keywords}: complex network; vital node; graph neural network; entropy
%\maketitle

\section{Introduction}
\label{sec1}
%% Labels are used to cross-reference an item using \ref command.

	In real-world networks, a few vital nodes often determine system behavior, particularly during intentional attacks or rumor propagation. Identifying these vital nodes remains a key challenge in complex network. 
	
	Many methods have been proposed, which can be roughly categorized into the neighbor-based centralities, path-based centralities and iteration-based centralities. A typical neighbor-based centrality is the degree centrality (DC)\cite{bonacich1972factoring}, which evaluates the node importance by calculating the nodes' degree. The $K$-shell\cite{kitsak2010identification} ranks the nodes by their positions in the network, which is also based on the nodes' degree. The path-based centralities include the betweenness centrality (BC)\cite{freeman1977set} and the closeness centrality (CC)\cite{freeman2002centrality}, etc. The betweenness centrality and the closeness centrality measure the node importance from the nodes' communication ability and the nodes' efficiency in information spreading, respectively. Iteration-based centralities include the eigenvector centrality (EC)\cite{bonacich1987power} and PageRank\cite{brin1998anatomy}, etc. The EC method considers the eigenvalue and eigenvectors of the adjacency matrix to measure the node importance, and the PageRank is initially used to evaluate the importance of web page by analyzing the links between web pages. Except for these classical methods, improved methods have also been proposed, by including other more information, e.g., the entropy\cite{wang2020identifying,10192527}.
	
 	Recently, the machine learning or deep learning methods have been developed to identify the vital nodes. As a representative of graph machine learning, the graph embedding methods have been employed to the vital node identification. For example, Lu et al propose a JTNMFR method by applying non-negative matrix factorization to the weighted adjacency matrix, and using the communicability and similarity matrices as the regularization terms\cite{ LU2024102384}. Xie et al\cite{xie2025mega} propose a meta-path embedding-based attention model, to solve suboptimal identification of vital nodes in heterogeneous networks. Ullah et al\cite{ullah2025finding} introduce a graph-embedding-based centrality by considering the nodes' interconnectedness to better capture the nodes' proximity. Lu et al\cite{lu2025mnegc} propose a network-embedded gravity model by using a multi-feature node mass and replacing the shortest path distance with the Node2vec-based distance. The entropy information is also introduced to combine with the network embedding to identify the vital nodes\cite{Lu_2023}.
	
	Graph Neural Networks (GNNs) learn the network features based on graph deep learning\cite{xu2018powerful,zhou2020graph,wu2020comprehensive}. The mainstream of the graph neural network includes the GraphSage\cite{hamilton2017inductive}, Graph Convolutional Networks (GCN)\cite{kipf2016semi}, and Graph Attention Network (GAT)\cite{velivckovic2017graph}. Zhao et al\cite{zhao2020infgcn} propose an InfGCN method, which overcomes the shortcoming of single perspective consideration in either the network structure or the node features in many methods. Bhattacharya et al\cite{bhattacharya2023detecting} propose a DeepInfNode method based on GCN, which can effectively identify the vital nodes in large-scale networks. Many methods suffer from insufficient generalization ability, and Liu et al\cite{liu2022learning} propose an improved method based on self-supervised learning and GCN, which adopts a multi-task learning framework and improves the generalization capability of traditional methods. Li et al\cite{li2024node} combine the GCN with a mini-batch training technique, integrating both the local and global structural information to improve the important node identification. Sun et al\cite{SUN2025102632} aggregate 7-hop neighbors and combine a Transformer encoder and Convolutional Neural Network to capture comprehensive information of nodes. Xiong et al\cite{xiong2024vital} introduce a AGNN method, which integrates Autoencoder and GNN to enhance node ranking performance in complex networks.
	
	Although many algorithms have been proposed, few methods have fully integrated the node feature, node interaction and node state. In this article, we propose a novel graph-neural-network-entropy (GNNE) method based on graph neural networks and information entropy, by jointly considering the above mentioned three aspects. The GNNE is composed of three parts, i.e., using a GCN to extract the network feature, a GAT to learn the node's influence factor, and the entropy to quantify the node's state in the network. We employ the GAT as the task learning model, which is trained on a Barabási–Albert (BA) network. Since the trained GAT requires feature input for testing on the real networks, we use the GCN to extract the network features, and the extracted features are then used as the input of the GAT model. However, as the GAT is not specifically designed for node importance ranking, we calculate the entropy of the node scores produced by the GAT, which is used to quantify the state of the node in the network and thus to measure the importance of the node. Our main contributions as follows.

1. The proposed method takes advantage of two graphical neural network structures, i.e., the GCN and GAT. Firstly, use the GCN to extract the node feature. Graph convolutional computation is effective in learning node features, and therefore exhibits advantages in feature extraction. Then, the node feature is input as the initial feature of the GAT network, which assigns different weights to the neigbhors with different importance based on the attention mechanism, and therefore can well capture the complex interaction between nodes. The output of the GAT can represent the influence factor of nodes. 

2. The nodes' influence factor obtained by the graph neural network is incorporated into an entropy model, and we propose a novel GNNE method. The entropy can quantify the state of a physical system, and using the nodes' entropy based on the nodes' influence factor can well assess the node importance.

3. The network fragility under intentional attack and the network propagation are investigated. Different from many existing methods trained on real datasets, we train the proposed method on an artificial BA network, and test the method on different real datasets. Therefore, the proposed method can be applied into various real systems, showing a nice generalized capability. Tested on six real datasets, the proposed method outperforms eight conventional methods and four graph-machine-learning-based methods.

\section{Problem definition}

An undirected and unweighted complex network can be described as $G{\rm{ }} = {\rm{ }}\left( {V,{\rm{ }}E} \right)$, with $V$ to be the node set and $E$ to be the edge set. The adjacency matrix is $A={a_{ij}}$. If there is an edge between the node $i$ and $j$, then the ${a_{ij}} = 1$; otherwise, ${a_{ij}} = 0$.

The objective of our study is to compute an importance score for each node based on the network structure, thereby identify the key nodes. Therefore, we propose the GNNE model, which consists of three modules, i.e., the feature extraction, task learning and importance list obtaining modules. The feature extraction module is composed of a two-layer GCN. The adjacency matrix $A \in {R^{(n \times n)}}$ and Laplacian matrix $L \in {R^{(n \times n)}}$ constructed from the graph data are used as the input of the feature extraction module. The output of the feature extraction module is ${H^{(1)}} \in {R^{(n \times 64)}}$, providing the node features as the input of the task learning module, which is composed of a two-layer GAT followed by a linear layer. The output of the GAT is ${H^{(2)}} \in {R^{(n \times 16)}}$, and of the linear layer produces the influence factor $\hat y \in {R^{(n \times 1)}}$. The influence factor ${{{\hat y}}}$ is input to the importance list obtaining module by calculating the node's entropy as its importance score.

\section{Proposed method}

\begin{figure}[htb]
	\centering
	\includegraphics[width=14cm]{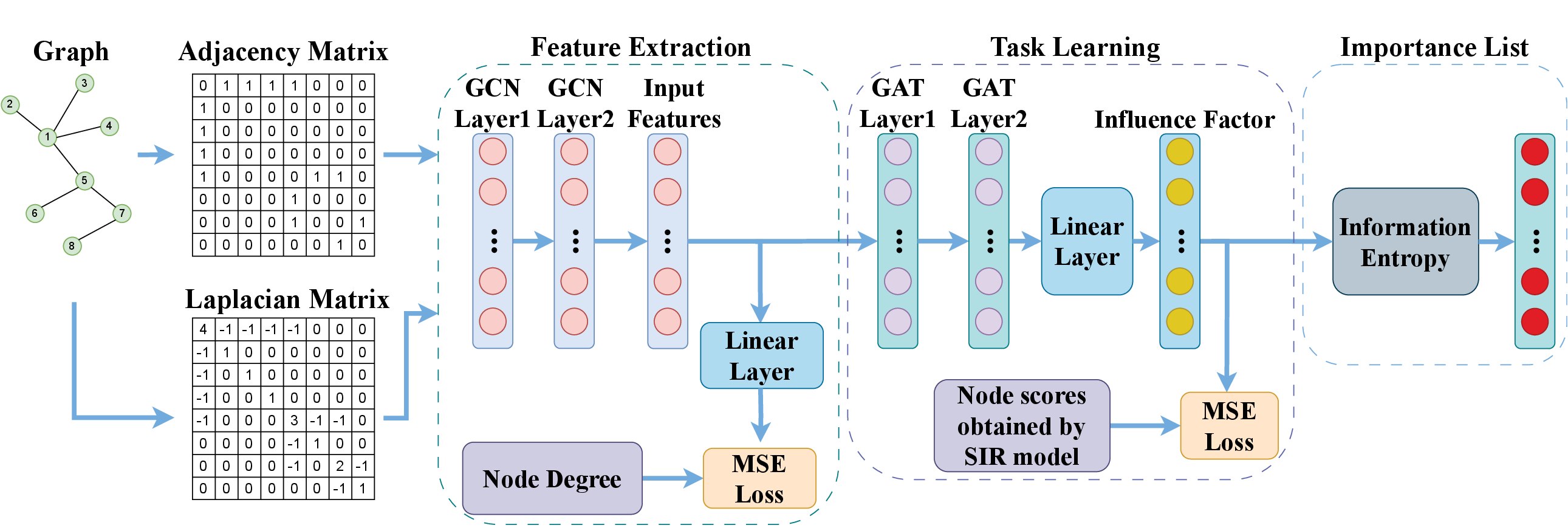}
	\caption{The workflow of the GNNE method. Firstly, the adjacency and Laplacian matrix are sent to the feature extraction module to generate the node features. Then, the node features are processed by the task learning module to output the influence factor of each node. Finally, the influence factor is input into the importance list obtaining module by calculating its entropy.}
\end{figure}

We propose the GNNE method by taking advantage of two graph neural network structures and information entropy. The main part of the GNNE include three modules, i.e., the feature extraction, task learning and importance list obtaining, with the workflow shown in figure 1. 

The feature extraction module is to extract the node features, which are used as the input of the task learning module. To improve the quality of feature extraction, we adopt two GCN layers and use the network topology feature measure, the node degree, as the labels for learning. The extracted features are then sent into the task learning module, which is based on two GAT layers. In many existing methods, the node importance scores obtained by the susceptible-infected-recovered (SIR) model are used as a benchmark. Here, we also use the SIR-based importance scores as the labels for training the task module. To enhance the model’s generalization ability, we train it on a synthetic BA network. The feature extraction and task learning modules are independent. Therefore, their training processes are separate. After the training process of the GAT, the node features of the real networks extracted by the GCN are then fed into the GAT to obtain the influence factor of the node. Finally, we calculate the entropy of each node based on its influence factor, which is used to assess the node's state and serves as an indicator of its importance in the network.

\subsection{Feature extraction}

We use the adjacency matrix of the network as the graph structure data and calculate the Laplacian matrix as the initial features , with the Laplacian matrix defined as: 

\begin{equation}
	L = D - A
\end{equation}
where $A$ is the adjacency matrix and $D$ is the degree matrix, with the $D$ to be a diagonal matrix whose elements on the diagonal are ${D_{ii}} = \sum\nolimits_j {{A_{ij}}} $. 

The adjacency matrix and Laplacian matrix are sent into the feature extraction module to obtain the feature. In our study, we use the graph convolutional network to extract the network feature. The feature extraction module is composed of two GCN layers and a linear layer, with the feature $H$ of the layer $l+1$ defined as,
\begin{equation}
	{{\rm{H}}^{{\rm{l}} + 1}} = \sigma ({D^{ - \frac{1}{2}}}\tilde A{D^{ - \frac{1}{2}}}{H^{(l)}}{W^{(l)}})
\end{equation}
where $\tilde A = A + I$, with $A$ to be the adjacency matrix and $I$ to be the identity matrix. $D$ is the degree matrix, ${W^l}$ is the weight matrix of the layer $l$ and $\sigma$ is the activation function.

Followed by the GCN is a linear layer,

\begin{equation}
	Y = XW + b
\end{equation}
where $X$ is the input feature vector. Then, we use the node degree as the label to calculate the loss,

\begin{equation}
	Loss1 = \frac{1}{N}\sum\limits_{i = 1}^N {{{({{\hat D}_i} - {D_i})}^2}} 
\end{equation}
where $N$ is the total number of nodes,  ${{\mathord{\buildrel{\lower3pt\hbox{$\scriptscriptstyle\frown$}} 
			\over D} }_i}$ is the model output of the node $i$ and ${D_i}$ is the degree of the node $i$. 

\subsection{Task learning}

The node feature vectors obtained from the feature extraction module are then sent into the task learning module, which consists of two GAT layers and one linear layer. The GAT utilizes the attention mechanism to assign different weights to each neighbor of a node, and can emphasize more on the neighbors with more importance. Therefore, the GAT is more advantageous in dealing with the network data with heterogeneity.

The computation of GAT has two steps. The first step is to calculate the attention coefficient ${e_{ij}}$,

\begin{equation}
	{{e_{ij}} = a([W{h_i}]\parallel [W{h_j}]),j \in {{\bar {\cal N}}_i}}
\end{equation}
where ${W}$ is the weight matrix, $\parallel $ denotes a join operation, $a()$ maps the feature to a real number, ${\bar {\cal N}_i} = {\cal N} \cup i$ is the set of the node $i$ and its neighboring nodes. Then, normalize the attention coefficient as,

\begin{equation}
	{{\alpha _{ij}} = \frac{{\exp (LeakyReLU({e_{ij}}))}}{{\sum\nolimits_{k \in {{\bar {\cal N}}_i}} {\exp (LeakyReLU({e_{ik}}))} }}}
\end{equation}

The second step is to do a weighted summation for the node features,

\begin{equation}
	h_i^{l + 1} = \sigma (\sum\limits_{j \in {{\bar {\cal N}}_i}} {{\alpha _{ij}}{W^{l + 1}}{h_j}^l} )
\end{equation}
where ${h_j}^l$ is the feature of the node $j$ in the layer $l$, ${W^{l + 1}}$ is the weight matrix of the layer $l+1$, ${\alpha _{ij}}$ is the normalized attention coefficient, and $\sigma $ is the activation function.

In order to enhance the stability of the model, we use the multi-head attention mechanism\cite{vaswani2017attention}, which is defined as follows,
\begin{equation}
	h_i^{l + 1} = \mathop \parallel \limits_{k = 1}^K \sigma (\sum\limits_{j \in {{\bar {\cal N}}_i}} {{\alpha _{ij}}^k{W^{l + 1}}{h_j}^l} )
\end{equation}
where ${\alpha _{ij}}^k$ is the normalized attention coefficient computed by the $k$-th attention mechanism.

Followed by the GAT is a linear layer,

\begin{equation}
	Y = XW + b
\end{equation}

The output of the linear layer is a real number, which represents the final extracted node feature, and therefore can be used as the influence factor of the node. 

We use the node scores simulated by the Susceptible–Infected–Recovered (SIR) model\cite{newman2002spread} as the label to calculate the loss. The SIR model is a well-known epidemic model, with the $S$ to be the susceptible individuals, $I$ to be the infected ones, and $R$ to be those recovered. The infected node would infect its susceptible neighbors with a probability $\beta $, and the infected nodes would recover with a probability $\gamma $. The threshold probability $\beta_{th}=\frac{{ < k > }}{{ < {k^2} >  -  < k > }}$ is reported to be very important in the SIR model\cite{lloyd2001viruses}, above which the infected individuals would much increased. Therefore, we use the average node scores of 1000-time SIR simulations at the threshold probability $\beta_{th}$ as the label to calculate the loss, 

\begin{equation}
	Loss2 = \frac{1}{N}\sum\limits_{i = 1}^N {{{({{\hat y}_i} - {y_i})}^2}} 
\end{equation}
where $N$ is the total number of nodes, ${{\hat y}_i}$ is the influence factor of the node $i$ and ${y_i}$ is the node score obtained by the SIR model.  

\subsection{Importance list basd on node entropy}
The entropy is widely used to quantify the state of a physical system, and we calculate the nodes' entropy based on the nodes' influence factor,

\begin{equation}
{E_i} =  - \sum\limits_{j \in {{\cal N}_i}} {\frac{{{y_i}}}{{{Y_j}}}{{\log }_2}} \frac{{{y_i}}}{{{Y_j}}}  
\end{equation}
where ${{\cal N}_i}$ is the neighbor set of the node $i$, $\frac{{{y_i}}}{{{Y_j}}}$ denotes the probability that the node $i$ is selected from its neighboring nodes, with ${y_i}$ to be the influence factor of the node $i$, ${Y_j} = \sum\limits_{k \in {{\cal N}_j}} {{y_k}} $, and ${{\cal N}_j}$ to be the neighbor set of the node $j$.

The entropy ${E_i}$ of the node $i$ quantifies its importance in the network. Therefore, we rank the entropy ${E}$ of the nodes, and the nodes with a high ${E}$ are identified as the vital nodes. The pseudo-code of the GNNE model is shown in algorithm 1.

\begin{breakablealgorithm}
	\caption{The pseudo-code of the GNNE }
	\label{alg:AOS}
	\renewcommand{\algorithmicrequire}{\textbf{Input:}}
	\renewcommand{\algorithmicensure}{\textbf{Output:}}
	
	\begin{algorithmic}[1]
		\REQUIRE Training BA network; Real-world network.  %%input
		\ENSURE The importance list of the real-world network.    %%output
		
		\STATE Calculate the adjacency matrix $A$ and the Laplacian matrix $L$ of the network;
		\STATE  GCN model parameter initialization;
		
		\FOR{t=1,2,…,${T_1}$ epoch}
		\STATE Input $A$ and $L$ into the GCN model to obtain the feature matrix $H$;
		\STATE $H$ is input into the linear layer to obtain the ${{\mathord{\buildrel{\lower3pt\hbox{$\scriptscriptstyle\frown$}} 
					\over D} }_i}$ of each node;
		\FOR{the node $i$ in the network}
		\STATE Calculate the loss between the degree ${D_i}$ and ${{\mathord{\buildrel{\lower3pt\hbox{$\scriptscriptstyle\frown$}} 
					\over D} }_i}$;
		\ENDFOR
		\STATE Backpropagation and parameter updates;
		\ENDFOR
		\STATE  Get the feature matrix of the training network ${H_{{\rm{train}}}}$ and the real-world network ${H_{{\rm{real}}}}$;
		\STATE  // Model training
		\FOR{the node $i$ in training network}
		\STATE Calculate the node score ${y_i}$ obtained by the SIR model;
		\ENDFOR
		\STATE GAT model parameter initialization;
		\FOR{t=1,2,…,${T_2}$ epoch}
		\STATE Feed ${A_{{\rm{train}}}}$ and ${H_{{\rm{train}}}}$ into the GAT model for attention mechanism learning;
		\STATE Get the influence factor ${{\hat y}_i}$ by a linear layer;
		\FOR{the node $i$ in the network}
		\STATE Calculate the loss between ${y_i}$ and ${{\hat y}_i}$;
		\ENDFOR
		\STATE Backpropagation and parameter updates;
		\ENDFOR
		\STATE  // Model testing
		\STATE  Feed ${A_{{\rm{real}}}}$ and ${H_{{\rm{real}}}}$ into the GAT model;
		\STATE  Get the influence factor of each node of the real network;
		\STATE  Calculate the information entropy based on the influence factor to obtain the node importance list.
		
	\end{algorithmic}
\end{breakablealgorithm}

\subsection{Complexity analysis}
The time complexity of the proposed GNNE is estimated as follows. For the feature extraction module, the time complexity of the GCN layer is $O(|E| \cdot d + N \cdot {d^2})$, where $N$ is the number of nodes, $|E|$ is the number of edges, and $d$ is the feature dimension, and of the linear layer is $O(N \cdot d)$. Therefore, the overall time complexity of the feature extraction module is $O(|E| \cdot d + N \cdot {d^2})$. For the task learning module, the time complexity of the GAT layer is $O(K \cdot |E| \cdot d + K \cdot N \cdot {d^2})$, where $K$ is the number of attention heads, and of the linear layer is $O(N \cdot d)$. Hence, the overall time complexity of the task learning module is $O(K \cdot |E| \cdot d + K \cdot N \cdot {d^2})$. For the importance list obtaining module, the time complexity is $O(|E| + N\log N)$. Therefore, the overall time complexity of the GNNE is $O(K \cdot |E| \cdot d + K \cdot N \cdot {d^2} + N\log N)$.

\section{Algorithms for comparison}
To comprehensively evaluate the proposed method, we compare it with an extensive baseline methods, including eight traditional topology-based methods and four graph-machine-learning-based methods. 

(1)	Degree Centrality (DC)\cite{bonacich1972factoring}: The degree centrality is defined as,
\begin{equation}
	D{C_i} = \frac{{{k_i}}}{{N - 1}},
\end{equation}
where ${k_i}$ denotes the degree of the node $i$ and $N$ is the total number of nodes.

(2)	$K$-shell\cite{kitsak2010identification}: The $K$-shell decomposition proceeds as follows: First, the nodes with a degree of $1$ are removed from the network, with their $K$-shell value $K_s$ to be 1. Second, the nodes with a degree of 2 are deleted from the network, with their $K$-shell value $K_s$ to be 2. This process is repeated until all nodes are assigned a $K_s$ value.

(3)	Betweenness Centrality (BC)\cite{freeman1977set}: The betweenness centrality is defined as,
\begin{equation}
	B{C_i} = \frac{{\sum\limits_{i \ne j \ne k}^N {{L_{jk(i)}}} }}{{\sum\limits_{j \ne k}^N {{L_{jk}}} }}
\end{equation}
where ${L_{jk}}$ is the number of shortest paths between the nodes $j$ and $k$, and ${L_{jk(i)}}$ is the number of shortest paths that pass through the node $i$.

(4)	Closeness Centrality (CC)\cite{freeman2002centrality}: The closeness centrality measures the average shortest distance from a node to all other nodes, which is defined as,
\begin{equation}
	C{C_i} = \frac{{N - 1}}{{\sum\nolimits_{j = 1}^N {{d_{ij}}} }}
\end{equation}
where ${d_{ij}}$ denotes the shortest path from the node $i$ to the node $j$, and $N$ is the total number of nodes.

(5)	Eigenvector Centrality (EC)\cite{bonacich1987power}: The eigenvector centrality is defined as,

\begin{equation}
	E{C_i} = {\lambda ^{ - 1}}\sum\limits_{j = 1}^N {{a_{ij}}{e_j}}, 
\end{equation}
where $\lambda $ is the maximum eigenvalue of the adjacency matrix $A$, and $e = {[{e_1},{e_2}, \cdots ,{e_N}]^T}$ is the eigenvector corresponding to the $\lambda $.

(6)	Harmonic Centrality (HC)\cite{boldi2014axioms}: Harmonic Centrality is a variant of the closeness centrality, which is defined as,
\begin{equation}
	H{C_i} = \frac{{\sum\nolimits_{j = 1}^N {\frac{1}{{{d_{ij}}}}} }}{{N - 1}}
\end{equation}

(7)	Collective Influence (CI)\cite{morone2015influence}: The collective influence centrality measures the collective influence by considering the neighbors belonging to the frontier $\partial Ball(i,\ell )$ of a ball, which is defined as,
\begin{equation}
	C{I_\ell }(i) = ({k_i} - 1)\sum\limits_{j \in \partial Ball(i,\ell )} {({k_j} - 1)} 
\end{equation}
where $\ell $ is the radius of a ball around each node.

(8) Improved $K$-shell (IKS)\cite{wang2020identifying} : The IKS is a hybrid centrality measure that combines the traditional $K$-shell decomposition with the node information entropy. For each node 
	$i$, the entropy is computed as,
	\begin{equation}
		{e_i} =  - \sum\limits_{j \in \Gamma (i)} {{I_j} \cdot \ln {I_j}} 
	\end{equation}
	where ${\Gamma (i)}$ denotes the neighbors of the node $i$, and ${I_j} = \frac{{{k_j}}}{{\sum\nolimits_{v = 1}^N {{k_v}} }}$  with the ${{k_j}}$ to be the degree of the node $j$ and $N$ to be the number of nodes in the network. The nodes are firstly grouped by $K$-shell, and then ranked according to the descending order of entropy in each shell.

(9) GAT\cite{velivckovic2017graph}: The method integrates two GAT layers followed by a linear output layer. Node features are initialized via the Node2Vec algorithm. The optimizer and other parameters, training data of the BA network, and node scores obtained by SIR are the same as those in the GNNE, to ensure a consistent evaluation. 

(10) GCN\cite{kipf2016semi}: The method integrates two GCN layers followed by a linear output layer. The initial node features, optimizer and other parameters are the same as the GAT method. 

(11) Autoencoder-GNN (AGNN)\cite{xiong2024vital}: The AGNN is a deep learning model that combines an Autoencoder and a GNN to obtain the node importance scores. The Autoencoder extracts structural features from the network, and the GNN predicts node importance scores.

(12) Graph-embedding-based hybrid centrality (GEHC)\cite{ullah2025finding}: The GEHC combines the DeepWalk-based graph embedding with hybrid centrality measures, i.e., the degree and $K$-shell centrality. The influence of the node $i$ is calculated as:
\begin{equation}
	GEHC(i) = \sum\limits_{j \in \eta (i)} {((w({v_i}) \times w({v_j})) \times {e^{ - |{r_i} - {r_j}{|^2}}})} 
\end{equation}
where $w({v_i}) = d({v_i}) \times ks({v_i})$, with $d({v_i})$ to be the degree centrality of the node ${v_i}$, and $ks({v_i})$ to be its $K$-shell value. ${|{r_i} - {r_j}|}$ is the Euclidean distance between the embedding vectors ${r_i}$ and ${r_j}$ obtained by the DeepWalk.

\section{Experiment}

\subsection{Datasets}

To enhance the generalization capability of the methods, we train the method on a synthetic BA network\cite{barabasi1999emergence}, containing 1000 nodes with an average degree of 4. Therefore, the training process is independent on any real network. Then, we use the trained model to test on six real networks\cite{rossi2015network}, with the data properties shown in Table 1. The six datasets are as follows: a United States air transportation network (USAir\cite{colizza2007reaction}), an email communication network (Email\cite{guimera2003self}), a network of American political weblogs (Polblogs\cite{adamic2005political}), 
a citation network of scientific papers (Cora\cite{sen2008collective}), a collaboration network in computational geometry(Geom\cite{rossi2015network}), and an electrical power network (Power\cite{watts1998collective}). 

\begin{table}[t]%% placement specifier
	\caption{Statistical properties of the six real networks. $N$ is the number of nodes, $M$ is the number of links, $\langle k\rangle$ is the average degree and $\langle C \rangle$ is the average clustering coefficient.}\label{fig1}
    \begin{tabular*}{\textwidth}{@{}l*{15}{@{\extracolsep{0pt plus
					12pt}}l}}
		%% Tabular cells are separated by &
		\toprule 
		Networks & $N$    & $M$     & $\langle k \rangle$  & $\langle C \rangle$ \\ 
		\midrule
		USAir  & 332 & 2126 & 12.8072  & 0.6252  \\ 
		Email  & 1133 & 5451  & 9.6222  & 0.2202   \\ 
		Polblogs  & 1222 & 16714 & 27.3552  & 0.3203  \\ 
		Cora  & 2485 & 5069  & 4.0797  & 0.2376  \\ 
		Geom  & 3621 & 9461 & 5.2256  & 0.5398  \\ 
		Power  & 4941 & 6594  & 2.6691  & 0.0801 \\ 
		\bottomrule
	\end{tabular*}
	%% Use \caption command for table caption and label.
\end{table}

\subsection{Metrics}
The method performance is evaluated from two perspective. One is from the network robustness under intentional attack. Specifically, if removing part of nodes would cause the network collapse, we then consider these nodes are the vital nodes of the network. Here we use two widely used metrics, i.e., the relative size of the largest connected component ($LCC$)\cite{morone2015influence} and the network efficiency\cite{latora2001efficient}. The other is from the spreading ability\cite{shi2023cost}. If some infected seed nodes would cause a wide range of infection, we then take these nodes as the important nodes of the network.

(1) The relative size of the largest connected component ($LCC$). The $LCC$ is defined as the ratio of the node number of the largest connected component after removing part of nodes to that of the original network,
\begin{equation}
	LCC = \frac{{{N_i}}}{N}
\end{equation}
where ${N_i}$ is the node number in the largest connected component after removing the node $i$, and $N$ is the node number of the original network.

(2) Network efficiency. The network efficiency of the network is defined as,
\begin{equation}
	\mu  = \frac{{\sum\limits_{i \ne j} {\frac{1}{{{d_{ij}}}}} }}{{N(N - 1)}}
\end{equation}
where ${d_{ij}}$ is the shortest distance between the node $i$ and $j$.

(3) Spreading ability. The spreading probability is simulated by the SIR model, with the infection probability $\beta $ to be the threshold probability, and the node recovery probability to be 1. Initially, we set the top 5\% nodes ranked by different methods as the infected nodes, and the rest nodes are susceptible nodes. The spreading ability $F(t)$ at time $t$ is defined as,
\begin{equation}
	F(t) = I(t) + R(t)
\end{equation}
where $I(t)$ and $R(t)$ are the infected and recovered nodes at time $t$, respectively.

\subsection{Parameter settings}
In the feature extraction module, the learning rate is set to be 0.001, the weight decay is set to be 0.0005, and the epoch is set to be 500. The first GCN layer has an input dimension equal to the number of nodes $n$ and an output dimension of 16. The second GCN layer has an input dimension of 16 and an output dimension of 64. 

In the task learning module, the learning rate and the weight decay are set to be the same as those of the feature extraction module, and the epoch is set to be 2000. The first GAT layer has an input dimension of 64 and an output dimension of 16, and the number of attention heads is 2. The second GAT layer has an input dimension of 32 and an output dimension of 16, and the number of attention heads is 1. The optimizer is the Adam optimizer for both the feature extraction and task learning modules.

\subsection{Results}

The relative size of the largest connected component $LCC$ on the node removal ratio $r$ is investigated, with the results shown in figure 2. It is observed that the $LCC$ of the GNNE decreases faster than all the baseline methods in all the datasets, and the advantages in the $LCC$ are more prominent in the Email and Polblogs datasets. At a critical point, the network would break down, with the $LCC$ approaching to 0. The critical points are found to be about $r= 0.24$ in the USAir dataset, $r=0.36$ in the Email dataset, $r=0.38$ in the Polblogs dataset, $r=0.18$ in the Cora dataset, $r=0.1$ in the Geom dataset, and $r=0.1$ in the Power dataset, which are found to be smaller than all other baseline methods. It suggests that the GNNE can well identify the vital nodes under intentional network attack. 

To better show the results, we then investigate how many nodes would be removed as the newtork is largely broken. Specifically, a smaller removal ratio leads to a small $LCC$, and the better is the algorithm. Here, we investigate the node removal ratio at around $LCC=0.01$, with the results shown in table 2. One may find that the GNNE significantly outperforms all the baseline methods in all the datasets except for the Cora dataset. In the Cora dataset, the removal ratio of the GNNE is slightly higher than that of the DC, but still much smaller than all the other baseline methods. It well supports the results shown in figure 2.

\begin{figure}[htb]
	\centering
	\includegraphics[width=14cm]{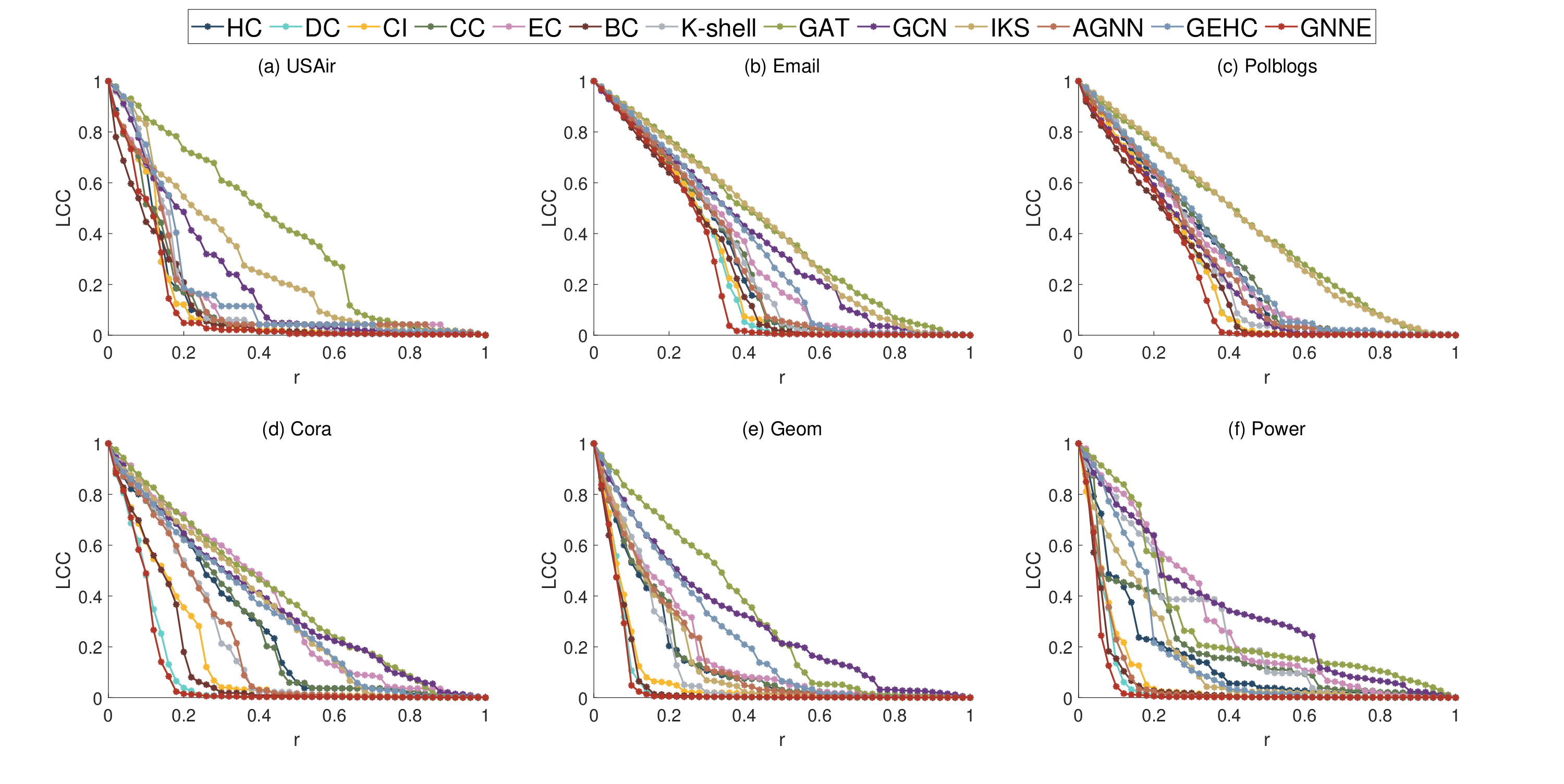}
	
	\caption{The relative size of the largest connected component $LCC$ on the node removal ratio $r$.}
\end{figure}

\begin{table}[t]%% placement specifier
	\caption{The ratio of removed nodes at around $LCC=0.01$, with the best results bolded.}\label{fig1}
	\begin{tabular*}{\textwidth}{@{}l*{15}{@{\extracolsep{0pt plus
						12pt}}l}}
		\toprule % 
		Methods & USAir    & Email     & Polblogs & Cora & Geom      & Power          \\
		\midrule % 
		HC  & 0.9789 & 0.7299 & 0.6817 & 0.9602 & 0.8937 & 0.9545  \\ 
		DC  & 0.6235 & 0.4766  & 0.4493 & \textbf{0.2419} & 0.1825 & 0.1973  \\ 
		CI  & 0.8102 & 0.5154 & 0.4673 & 0.5960 & 0.5123 & 0.3206   \\ 
		CC  & 0.9669 & 0.7449  & 0.6866 & 0.9646 & 0.9116 & 0.9842  \\ 
		EC  & 0.9669 & 0.7996 & 0.6612 & 0.9606 & 0.9254 & 0.8298  \\ 
		BC  & 0.5392 & 0.5463 & 0.4427 & 0.3871 & 0.1969 & 0.3507  \\ 
		$K$-shell  & 0.8223 & 0.5755  & 0.5786 & 0.5553 & 0.3598 & 0.6646   \\ 
		GAT  & 0.9428 & 0.9312  & 0.9198 & 0.9296 & 0.8108 & 0.9759   \\ 
		GCN  & 0.7892 & 0.8605  & 0.5720 & 0.9557 & 0.9652 & 0.9557   \\ 
		IKS  & 0.9518 & 0.8694  & 0.9313 & 0.7054 & 0.5744 & 0.7146   \\ 
		AGNN  & 0.9729 & 0.6231  & 0.6514 & 0.5976 & 0.6377 & 0.2613   \\ 
		GEHC  & 0.9398 & 0.6681  & 0.7938 & 0.7944 & 0.6868 & 0.5847   \\ 
		GNNE  & \textbf{0.4789} & \textbf{0.4528}  & \textbf{0.3977} & 0.2423 & \textbf{0.1428} & \textbf{0.1730}   \\ 
		\bottomrule 
	\end{tabular*}
	%% Use \caption command for table caption and label.
\end{table}

The network efficiency $\mu$ on the node removal ratio $r$ is show in figure 3. It is observed that the network efficiency $\mu$ decays fastest with the node removal ratio $r$ in the GNNE for almost all the datasets, especially in the Email and Polblogs datasets. The critical points of the network fragmentation are at about  $r= 0.2$ in the USAir dataset, $r=0.36$ in the Email dataset, $r=0.34$ in the Polblogs dataset, $r=0.14$ in the Cora dataset, $r=0.1$ in the Geom dataset and $r=0.1$ in the Power dataset, smaller than all the other baseline algorithms. Similarly, we also compute the node removal ratio at a small network efficiency around $\mu  = 0.01{\mu _0}$, where $\mu _0$ is the initial network efficiency. The results are shown in table 3. Also, we can find that the GNNE presents a much smaller node removal ratio, i.e., removing relatively fewer nodes would greatly damage the network structure in the GNNE. That is, the network efficiency results also manifest that the GNNE can better rank the nodes with different importance. 

\begin{figure}[htb]
	\centering
	\includegraphics[width=14cm]{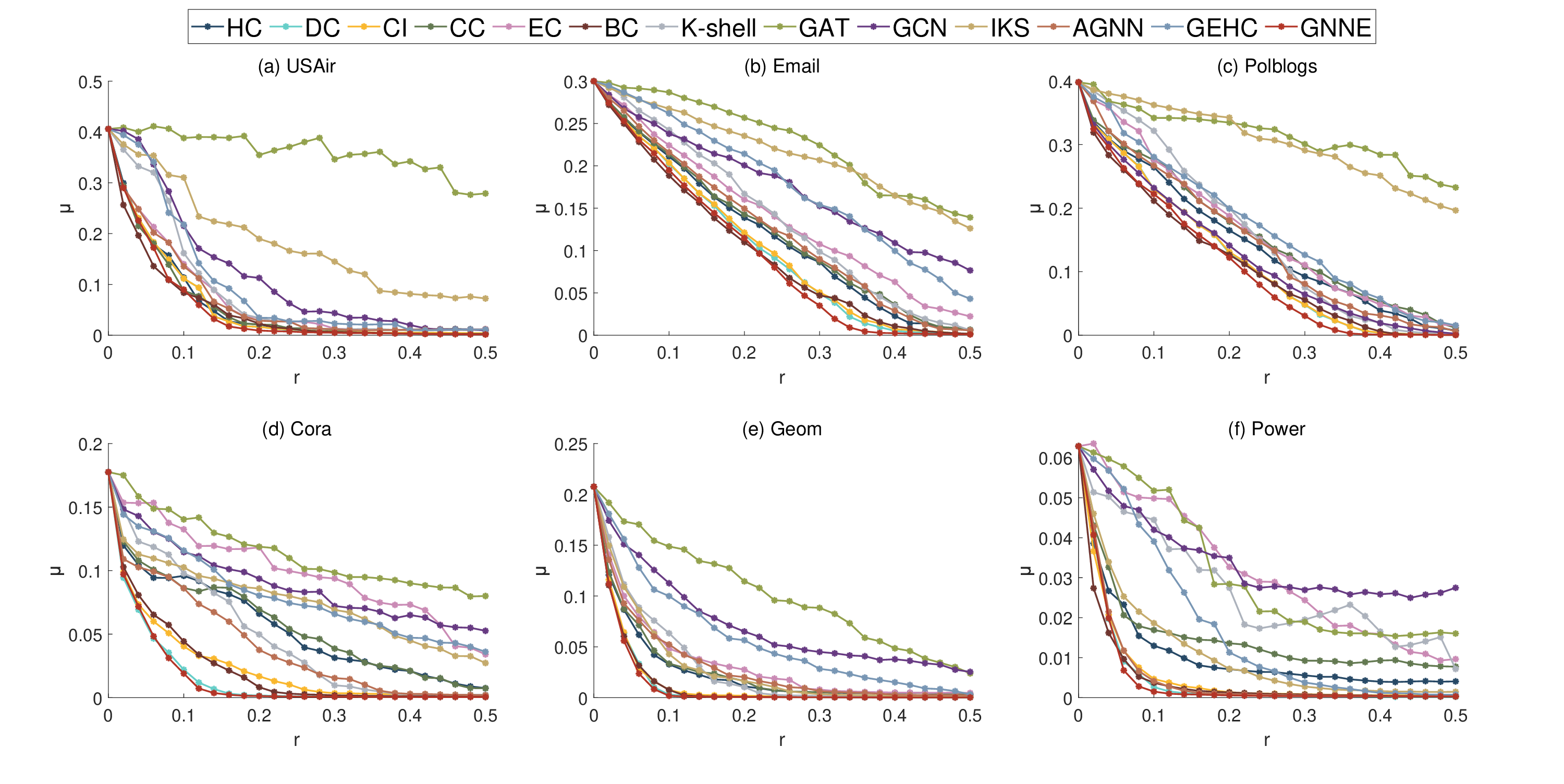}
	
	\caption{The network efficiency $\mu$ on the node removal ratio $r$.}
\end{figure} 

\begin{table}[t]%% placement specifier
	\caption{The ratio of removed nodes at around $\mu  = 0.01{\mu _0}$, with the best results bolded.}\label{fig1}
	\begin{tabular*}{\textwidth}{@{}l*{15}{@{\extracolsep{0pt plus
						12pt}}l}}
		\toprule % 
		Methods & USAir    & Email     & Polblogs & Cora & Geom      & Power          \\ 
		\midrule % 
		HC  & 0.9789 & 0.6399 & 0.5139 & 0.9996 & 0.9992 & 0.9986  \\ 
		DC  & 0.4669 & 0.4431  & 0.3944 & 0.1895 & 0.1141 & 0.1973  \\
		CI  & 0.4669 & 0.4704 & 0.3912 & 0.5082 & 0.1853 & 0.3781   \\ 
		CC  & 0.9819 & 0.6593  & 0.5327 & 0.9996 & 0.9992 & 0.9992  \\ 
		EC  & 0.9819 & 0.7017 & 0.6064 & 0.9996 & 0.9992 & 0.9982  \\ 
		BC  & 0.3645 & 0.4766 & 0.4075 & 0.3111 & 0.1215 & 0.3447  \\ 
		$K$-shell  & 0.7108 & 0.5384  & 0.4501 & 0.5561 & 0.2684 & 0.6780   \\ 
		GAT  & 0.9940 & 0.9974  & 0.9992 & 0.9988 & 0.9997 & 0.9992   \\ 
		GCN  & 0.9247 & 0.9894  & 0.4877 & 0.9996 & 0.9997 & 0.9998   \\
		IKS  & 0.9578 & 0.9347  & 0.9755 & 0.9988 & 0.2793 & 0.9982   \\ 
		AGNN  & 0.9729 & 0.5808  & 0.5417 & 0.5400 & 0.5131 & 0.2629   \\ 
		GEHC  & 0.9639 & 0.6814  & 0.6334 & 0.9630 & 0.5675 & 0.5612   \\  
		GNNE  & \textbf{0.3524} & \textbf{0.3769}  & \textbf{0.3527} & \textbf{0.1706} & \textbf{0.0975} & \textbf{0.1864}   \\ 
		\bottomrule % 
	\end{tabular*}
	%% Use \caption command for table caption and label.
\end{table}

Figure 4 displays the spreading ability for different methods. The spreading ability $F(t)$ is observed to firstly increase with the evolution time $t$, and then reach steady. The more infected and recovered individuals at the steady state, the higher spreading ability of the vital nodes, i.e., the better the method. From the figure, the GNNE outperforms all the baseline methods in the USAir, Email, Polblogs, Cora, Geom networks. In the Power network, the GNNE is the second best, and its $F(t)$ is slightly smaller than that of the DC, but is much higher than those of other baseline methods. It demonstrates that the GNNE is also more advantageous in the spreading ability.

Overall, compared with eight traditional topology-based methods and four graph-machine-learning methods, the proposed GNNE is more advantageous in the relative size of the largest connected component, network efficiency, and spreading ability. 

\begin{figure}[htb]
	\centering
	\includegraphics[width=14cm]{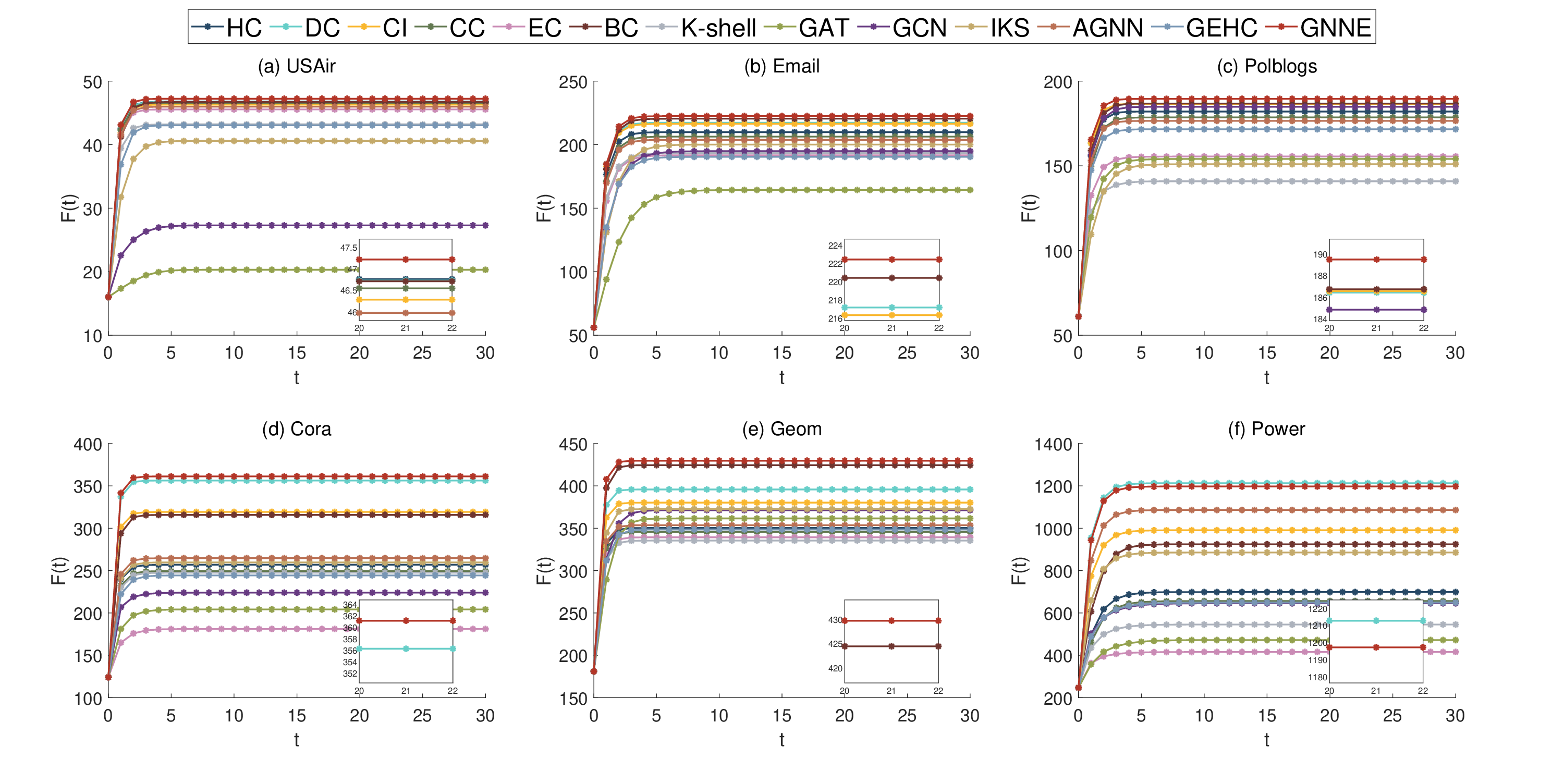}
	\caption{Spreading ability $F(t)$ on the evolution time $t$. The inner panels are the zoomed-in plots at the steady state.}
\end{figure}

\subsection{Parameter sensitivity analysis}
In graph neural networks, hyperparameters may influence the algorithm performance. Therefore, we conduct a parameter sensitivity analysis, focusing on three parameters, i.e., the layer number $L$ of the GCN and GAT networks, the feature dimension $d$ generated by the feature extraction module, and the head number $K$ of the multi-head attention mechanism in the GAT network. When adjusting one parameter, other parameters keep fixed. Figure 5 presents the results of the relative size of the largest connected component $LCC$, network efficiency $\mu$, and spreading ability $F(t)$ under different parameter values. Here, the $LCC$ and network efficiency $\mu$ are calculated when the nodes are removed up to the critical points of each dataset mentioned in section 5.4, and $F(t)$ is calculated at the steady state. The results indicate that the spreading ability remains highly stable for different parameter values. The $LCC$ and the network efficiency $\mu$ are also stable in most datasets, but fluctuate in few cases. For example, the Email dataset shows a relatively large value at $d$=128. From the results, the algorithm generally achieves an optimal performance at the layer number $L=2$, the feature dimension $d=64$, and the head number $K=2$.

\begin{figure}[htb]
	\centering
	\includegraphics[width=14cm]{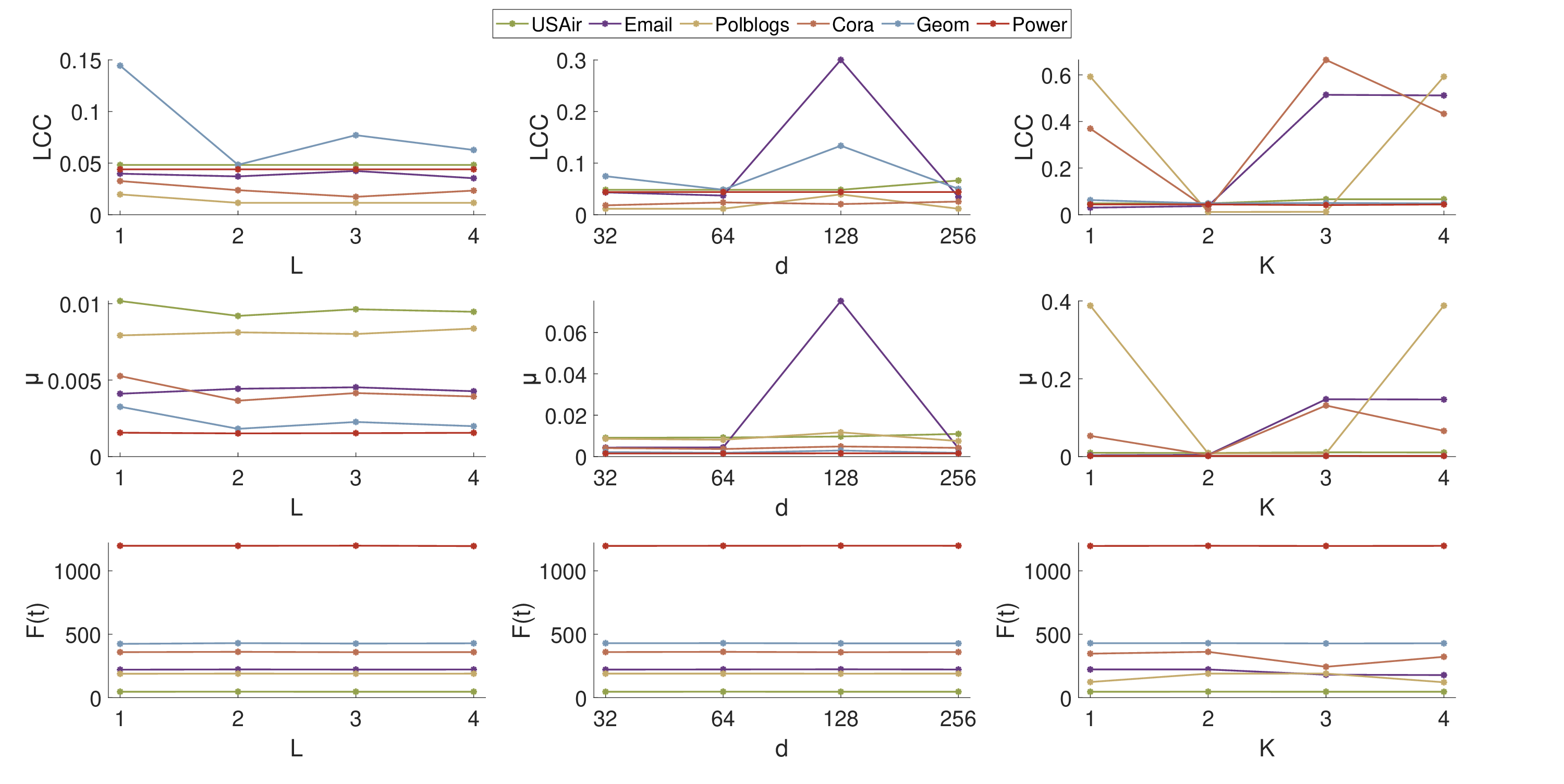}
	\caption{The $LCC$, $\mu$, and $F(t)$ on the GCN and GAT layer number $L$, feature dimension $d$, and head number $K$ of the attention mechanism in the GAT network. The $L=1, 2, 3,4$ indicates the layer number of both the GCN and GAT network is 1, 2, 3 and 4, respectively.}
\end{figure}

\subsection{Discussions}

As is well known, most real systems show scale-free characteristics, exhibiting strong heterogeneity. Training the proposed method on the BA network can therefore improve the generalization ability of the model. To show how the model is adaptive to different network structures, we investigate the structural characteristics of the six test networks. Figure 6 displays their degree distributions, and one may find that they also exhibit power-law distribution, however, their power-law exponents are significantly different. It indicates that the six datasets show different levels of heterogeneity. Moreover, as shown in Table 1, the average clustering coefficients of the six networks are also substantially different, reflecting diverse local connectivity patterns.

As two typical examples, the USAir and Power datasets demonstrate a contrasting structural characteristics. The USAir network has an average clustering coefficient of 0.6252, suggesting a highly clustered structure, and a degree distribution exponent around 0.95, indicating strong heterogeneity. In contrast, the Power network has an average clustering coefficient of 0.0801, suggesting a lowly clustered structure, and a degree distribution exponent around 3.95, implying a relatively weaker heterogeneity. Despite their differences in the structural characteristics, the proposed method achieves satisfactory results on both datasets, which demonstrates its favorable generalization ability for networks with distinct structural properties.

\begin{figure}[htb]
	\centering
	\includegraphics[width=14cm]{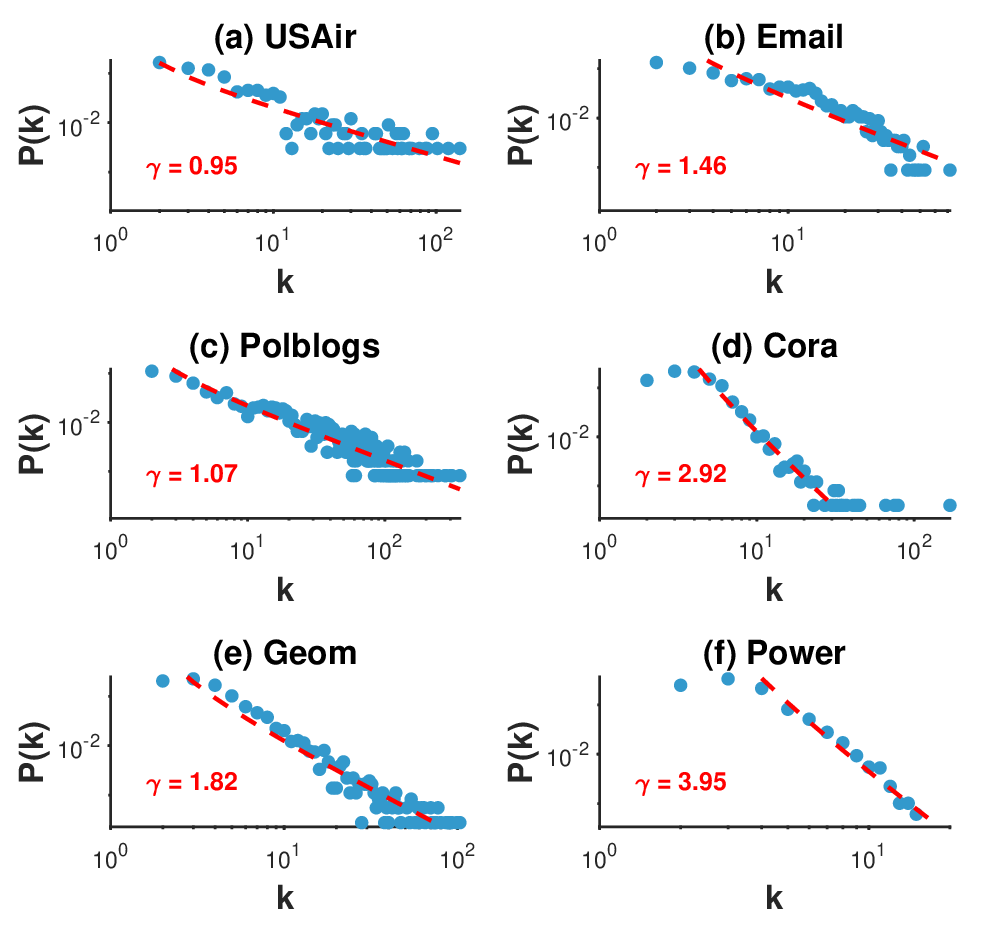}
	\caption{Degree distribution of the six datasets. $\gamma $ is the exponent of the power-law fitting.}
\end{figure}

\section{Conclusions}

We propose a GNNE method to identify the vital nodes of complex networks, based on the graph neural networks and information entropy. We first use the GCN to generate the nodes' features, then employ the GAT to capture the interaction features between the nodes and their neighbors, obtaining the nodes' influence factors, and finally calculate the nodes' entropy based on the influence factors to assess the node importance. The model is trained on the BA network and tested on six real datasets. The algorithm performance is evaluated from two perspectives of the network fragility under attack and the network propagation. Experimental results show that the proposed GNNE outperforms eight traditional topology-based methods and four graph-machine-learning-based methods in most cases.  

Many existing methods consider the network feature from some specific perspective, but cannot fully reflect the complex interaction between nodes and the nodes' state. The GNNE takes advantage of the GCN and GAT, which not only extracts the nodes' feature, but also considers the different importance of the neighboring nodes. The proposed method also employs the entropy to quantify the nodes' state, and therefore can better capture the network characteristics. Moreover, the proposed method is trained on the synthetic BA network, and therefore presents a nice generalized capability. Our study provides a promising way to the vital node identification from the perspective of fully extracting the network features.

{\textbf{Acknowledgments:}}
This work was partially supported by the National Natural Science Foundation of China (Grant No. 11865009).

\bibliographystyle{elsarticle-num}
\bibliography{reference}
\end{document}